\documentclass[%
 reprint,
 amsmath,amssymb,
 aps, superscriptaddress
]{revtex4-1}

\usepackage{graphicx}
\usepackage{dcolumn}
\usepackage{bm}
\usepackage{svg}

\usepackage{mathtools}
\usepackage{amsmath}%
\usepackage{amsfonts}%
\usepackage{amssymb}

\DeclarePairedDelimiter\ket{\lvert}{\rangle}
\DeclarePairedDelimiterX\braket[2]{\langle}{\rangle}{#1 \delimsize\vert #2}

\title[Probing photon statistics by means of field scattering onto the superconducting qubit]{The quantum and classical field scattered on a single two-level system}
\begin{document}

\preprint{APS/123-QED}

\title[Probing photon statistics by means of field scattering onto the superconducting qubit]{The quantum and classical field scattered on a single two-level system}

\author{S.A. Gunin}
\email{gunin.sa@phystech.edu}
\affiliation{Moscow Institute of Physics and Technology, Dolgoprudniy, Russia}
\affiliation{Skolkovo Institute of Technology, Moscow, Russia}
\author{A.Yu. Dmitriev}
\affiliation{Moscow Institute of Physics and Technology, Dolgoprudniy, Russia}
\affiliation{Russian Quantum Center, Skolkovo village, Russia}
\author{A.V. Vasenin}
\affiliation{Moscow Institute of Physics and Technology, Dolgoprudniy, Russia}
\affiliation{Skolkovo Institute of Technology, Moscow, Russia}
\author{K.S. Tikhonov}%
\affiliation{Moscow Institute of Physics and Technology, Dolgoprudniy, Russia}
\affiliation{Landau Institute for Theoretical Physics, Moscow 119334, Russia}
\author{G.P. Fedorov}%
\affiliation{Moscow Institute of Physics and Technology, Dolgoprudniy, Russia}
\affiliation{Russian Quantum Center, Skolkovo village, Russia}
\affiliation{National University of Science and Technology, MISIS, Moscow, Russia}%
\author{O.V. Astafiev}%
\affiliation{Skolkovo Institute of Technology, Moscow, Russia}
\affiliation{Moscow Institute of Physics and Technology, Dolgoprudniy, Russia}

\date{\today}
\begin{abstract}
In many problems, the scattering amplitudes of weak coherent pulse are almost equivalent to the ones of single propagating photon. We thoroughly compare the scattering of: (i) short microwave coherent pulse from rf generator and (ii) vacuum-photon coherent superposition from the two-level Emitter, both directed to a single two-level system – the Probe. To do that, we use two superconducting qubits to implement  Emitter and Probe, both strongly coupled to the same waveguide. However, with the use of magnetic circulator we couple the field from Emitter to the Probe without reverse backaction, thereby working with a cascaded atomic system implemented in waveguide-QED setup. By measuring the dynamics of scattered field, we find a certain discrepancy between two cases, confirmed by analytical and numerical study. Particularly, we find the optimal amplitude $\Omega_*$ of classical pulse mimicking the superposition from Emitter, for which the difference becomes very small (but non-vanishing), and is almost unavailable to measure in practice. 
\end{abstract}

\maketitle

\section{\label{sec:level1}Introduction}
An artificial superconducting atom is a promising platform for studying fundamentals of light-matter interaction due to the ability to achieve strong coupling with an open space or single-mode fields \cite{Wallraff, StrongCoupling}. Over the last two decades many quantum optical experiments demonstrating various properties of light were conducted: resonance fluorescence of the single qubit \cite{Astafiev2010}, wave mixing \cite{Dmitriev2017}, single-photon routing \cite{Delsing}.
Moreover, since the cQED platform provides a highly efficient on-demand single-photon source \cite{Peng} it is possible to study scattering of various light states on atoms strongly coupled to an open 1D space. Studying the interaction of light having arbitrary statistics with the single atom has a long-term theoretical interest \cite{Gardiner1994, Pogosov,  Liao, Fan2014}.

In the present work, we study both theoretically and experimentally scattering of the single-photon pulses being in the zero-one superposed photon state, hereinafter it is called quantum pulse, on the superconducting transmon qubit, and subsequently compare obtained results with the case of scattering exponentially modulated coherent pulse, henceforward we call it classical. The quantum pulse generated by the single-photon source lacks states $|n\rangle$ with $n>1$. Therefore, the scattering of such a pulse would not result in more than one atomic Rabi oscillation, which, in general, is not the case for the coherent pulse, which forces the atomic population to oscillate. Population oscillations are connected with the scattered field oscillations, specifically, with the field envelope. Measuring the time-domain envelope of the scattered field, we find that, in the limit of small field amplitude of the coherent signal, the experiment shows practically indistinguishable dynamics of the scattered fields, compared to the scattering of anti-bunched light, due to the relatively small impact of the higher order Fock states present in the coherent pulse.

However, theoretical analysis proves that the scattered field dynamics differs for all coherent pulse parameters, which is due to the third-order field corrections. Furthermore, we find the coherent field amplitude allowing one to find the closest classical approximation of the scattered field dynamics compared to the quantum case. Hence, in the low-signal limit, it allows one to consider single-photon approximation in a wide range of problems, as in the single-photon microwave range detector \cite{Inomata2016single}.

\section{Experimental results}

\subsection{\label{sec:level2}The device and measurement setup}
\begin{figure}[h]
\includegraphics[width=0.95\linewidth]{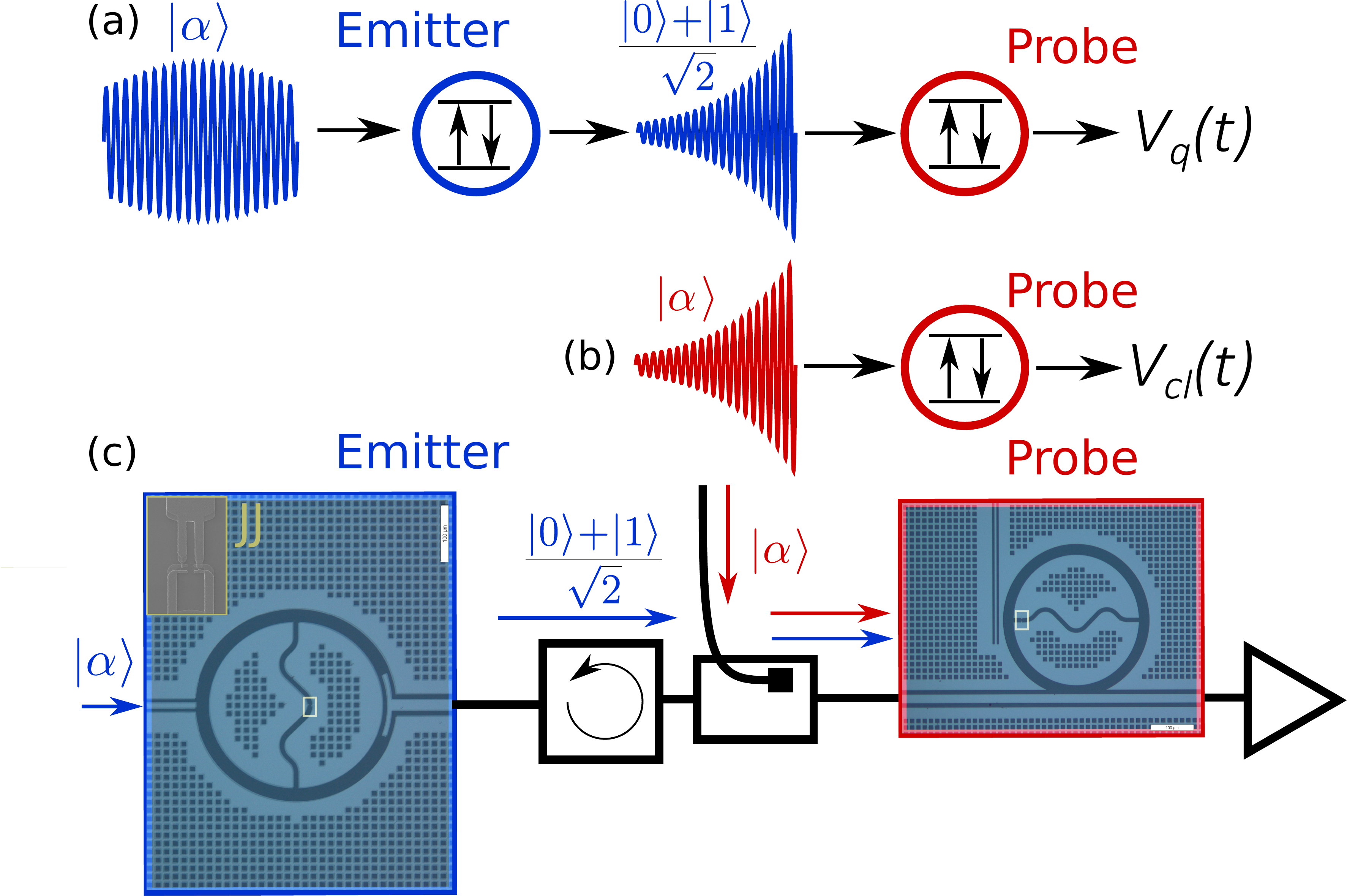}
\caption{\label{fig:Scheme2_v2} (a) Schematic problem statement for the scattering of the quantum signal being produced from Emitter qubit on the Probe qubit (b) Schematic problem statement for the scattering of the coherent signal on the Probe qubit. (c)  Schematic describing the connection between sample holders. Blue lines denote the signal path in scattering of the single-photon pulses, red lines denote the signal path in scattering of the coherent pulses. Output amplification line is common for both experiments. Optical images of Emitter-qubit and Probe-qubit are presented. The yellow box on (c) contains a scanning electron micrograph of the Josephson junctions}
\end{figure}

To explore the stated scattering problem, we implement a cascaded \cite{Roch2014, Fan2014} system from two superconducting artificial atoms --- tunable transmon qubits  \cite{Koch}, see Fig.~\ref{fig:Scheme2_v2}.
The first atom is coupled to a pair of semi-infinite waveguides, implementing a single-photon source (which we will call Emitter) \cite{Peng}.
It may be excited by a driving pulse applied via the weakly coupled line, preparing an arbitrary single-photon superposed state, and subsequently it radiates into the strongly coupled emission line.
The field emitted to the output waveguide propagates through a circulator and then interacts with the Probe (quantum scatterer) qubit, which is side-coupled to the waveguide. We subsequently detect the voltage in this waveguide $V_{q}(t)$, see Fig. 1(a). The circulator prevents the back-action of the Probe qubit on the Emitter qubit \cite{Gardiner1985, Gardiner1994}. We estimate power loss between the output of the Emitter and the coupling point of the Probe atom to be about 3 dB.

Emitter and Probe are located on two different silicon chips and placed into two separate sample holders with a separate permalloy magnetic shields. Each sample holder has its own magnetic coil, which allows one to change magnetic fluxes thereby tuning frequencies of the devices. The output signal from the Probe is then amplified by cryogenic and room-temperature amplifiers. After amplification, the signal is down-converted by a standard heterodyne scheme with quadrature mixers and both quadratures are digitized with fast ADCs. Moreover, due to the possibility of independent flux tuning, it is possible to detect scattered field part separately from an incoming pulse, since it can be independently recorded and subsequently substracted from  the experimental traces.

A coherent signal may also be applied directly to the Probe qubit via a directional coupler without disturbing the Emitter, the field scattered in this experiment we denote as $V_{cl}(t)$, see Fig. 1(b). 
Therefore, the presented setup allows one to study the scattering of either equal single-photon superposition ($\frac{|0\rangle+|1\rangle}{\sqrt{2}}$ state) or classical pulses on the Probe qubit, since the output line is common, see Fig. 1(c). 

\subsection{\label{sec:level2}Device characterization}
\begin{figure}[h]
\includegraphics[width=1\linewidth]{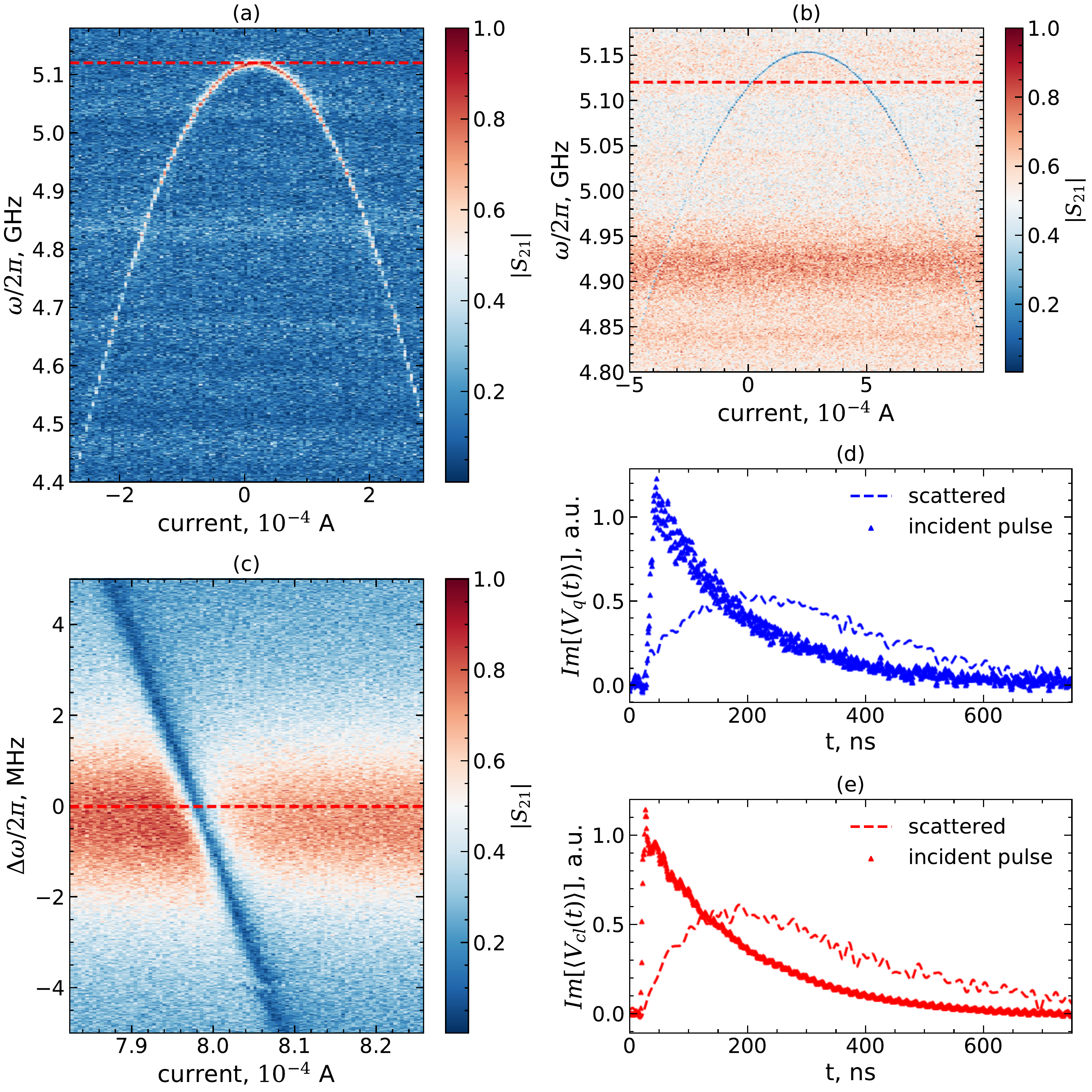}
\caption{\label{fig:spectra} (a) Single-tone spectroscopy of the bare Emitter, red dashed line represents sweet-spot location $\omega_{c}$. The emission line of the Probe is represented at the (b), red dashed line is located at $\omega=\omega_{c}$. When qubits are tuned into resonance the transmission dip occurs at the $\omega=\omega_{c}$ (c), the Probe is tuned throughout the Emitter's resonance, the Emitter is located at sweet-spot frequency $\omega_{c}$, detuning $\Delta \omega$ between scanning frequency $\omega$ and Emitter sweet-spot $\omega_{c}$ is defined as $\Delta \omega = \omega - \omega_{c}$. Figure (d): measurements with the quantum driving pulse emitted from the Emitter being resonant to Emitter with the Probe either in or out of resonance. The scattered field is represented with blue triangles, the exciting single-photon pulse is a blue dashed curve. Figure (e):
 The coherent pulse with the exponential envelope excites the Probe at $omega_{c}$ frequency. Incident exponential pulse is plotted as red dashed line. The envelope of scattered field on the Probe is represented with red triangles.}
\end{figure}
Firstly, we characterize energy level structure of the discussed devices.
Hence, we provide elastic scattering data of continuous waves for Emitter and Probe qubits. Fig.~\ref{fig:spectra}(a) presents the spectrum of Emitter versus bias current in a coil, generating a magnetic field, which penetrates the SQUID loop. The sweet-spot position of Emitter is located at $\omega_{c}/2\pi=5.119$~GHz, see dashed red line in Fig. 2(a).

Similarly, we record the bare spectrum of the Probe qubit, see Fig 2(b). The Emitter's sweet-spot slightly differs from the Probe's one $\omega_{p}/2\pi=5.153$~GHz. Therefore, in order to study resonant scattering we should fix the frequency $\omega_{c}$ and precisely tune Probe into resonance with $\omega_{c}$. As soon as the output of the Emitter at the frequency $\omega_{c}$ passes through the waveguide with the Probe being tuned to the $\omega_{c}$, one can observe the transmission dip from the Probe on the emission profile of the Emitter when two qubits are in resonance. And the Emitter peak when they are detuned, see Fig. 2(c). Further we tune resonant transmission dip location to the Emitter sweet-spot position. Hereinafter, this frequency would be a carrier for the subsequent experiments.

Next we proceed to time-resolved measurements, in order to characterize relaxation rates. We tune the Emitter into resonance with the carrier frequency. Radiative linewidth  of the Emitter equals $\Gamma/2 \pi=1.86$~MHz. Probe relaxation rate in that point is $\gamma/2\pi\simeq 1.85$~MHz.

Knowing relaxation rates, we are able to study resonantly scattered fields either after interaction of the Probe with the anti-bunched light, being generated from Emitter qubit, or with the coherent pulse of the same width independently scattered on the Probe. We generate equal vacuum-photon superposition $(|0\rangle+|1\rangle)/\sqrt{2}$ by sending $8$~ns long $\pi/2$ pulse in the control line of the Emitter. The averaged emitted field is the exponentially decaying wave packet, see blue dashed line Fig. 2(d). If we tune the Probe into the resonance with Emitter (and also with the driving pulse), what we observe is the exponential pulse interference with the field scattered by a Probe qubit, see blue triangles Fig. 2(d). In the case of coherent pulse scattering we apply exponentially modulated pulses and measure the average transmitted field through the Probe being either in resonance with the carrier or out of it. The original pulse transmitted through the waveguide is recorded when Probe is far detuned, see red triangles Fig. 2(e). If the Probe is in resonance with the carrier, see the red dashed line Fig. 2(e), we detect the superposition of the scattered and incident fields. The similarities and differences between resonantly scattered fields in the two described experiments are analyzed below.



\subsection{\label{sec:level2}Single-photon pulse scattering}

Firstly, we experimentally study details of the single-photon scattering case. We record the field envelopes $V_{q}(t)$, see Fig. 1(a), with respect to detuning $\Delta=\omega_{c}-\omega$ between single-photon carrier frequency $\omega_{c}$, which always remains constant, and Probe frequency $\omega_{p}$. Probe coil current is tuned non-linearly, using recorded earlier spectra, in order to achieve linear slope in Probe frequency. Detected field envelopes $V_{q}(t)$ with respect to detuning contain the incoming single-photon pulse and the scattered part, they are presented in the left column of Fig. 3. Since the number of photons in the quantum pulse is not greater than one, and consequently, if the Probe is in exact resonance with the quantum pulse, the envelope of scattered signal does not behave oscillatory, which is a certain signature of the photon statistics, since this would require more than one photon. Instead, it has one extremum and then decays to zero. So, beats of the envelope with non-zero detuning are not connected with the multi-photon processes. Also we see that fields are symmetric and anti-symmetric with respect to detuning sign, so in resonance condition after subtraction of the incoming pulse we should obtain zero quadrature.

Next we analyze the problem analytically. The field $V_{e}(t)$ emitted by the Emitter into the waveguide with impedance $Z_e$ at the frequency $\omega_{c} $is proportional to the expectation value of lowering operator for the Emitter $\langle\sigma^e_{-}\rangle$. 
\begin{equation}
    V_{e}(t) = A_e\sqrt{\gamma}\langle{\sigma^e_-(t)}\rangle,
    \label{eq: emission}
\end{equation}
where the constant $A_e = i\sqrt{\hbar\omega_c Z_e}$ is the same for any field scattered by Probe, and further it will be omitted. For typical parameters, the field amplitude is approximately $10^{-8}$ V before amplification. This field amplitude has no adjustable parameters, and the envelope shape is completely defined by the state of the Emitter. For the equal single-photon superposition we have $\langle\sigma^e_{-}\rangle(t) = \frac{\exp(-\gamma_2 t)}{2}$, where $\gamma_{2}$ is the dephasing rate.

\begin{figure}[h]
\includegraphics[width=\linewidth]{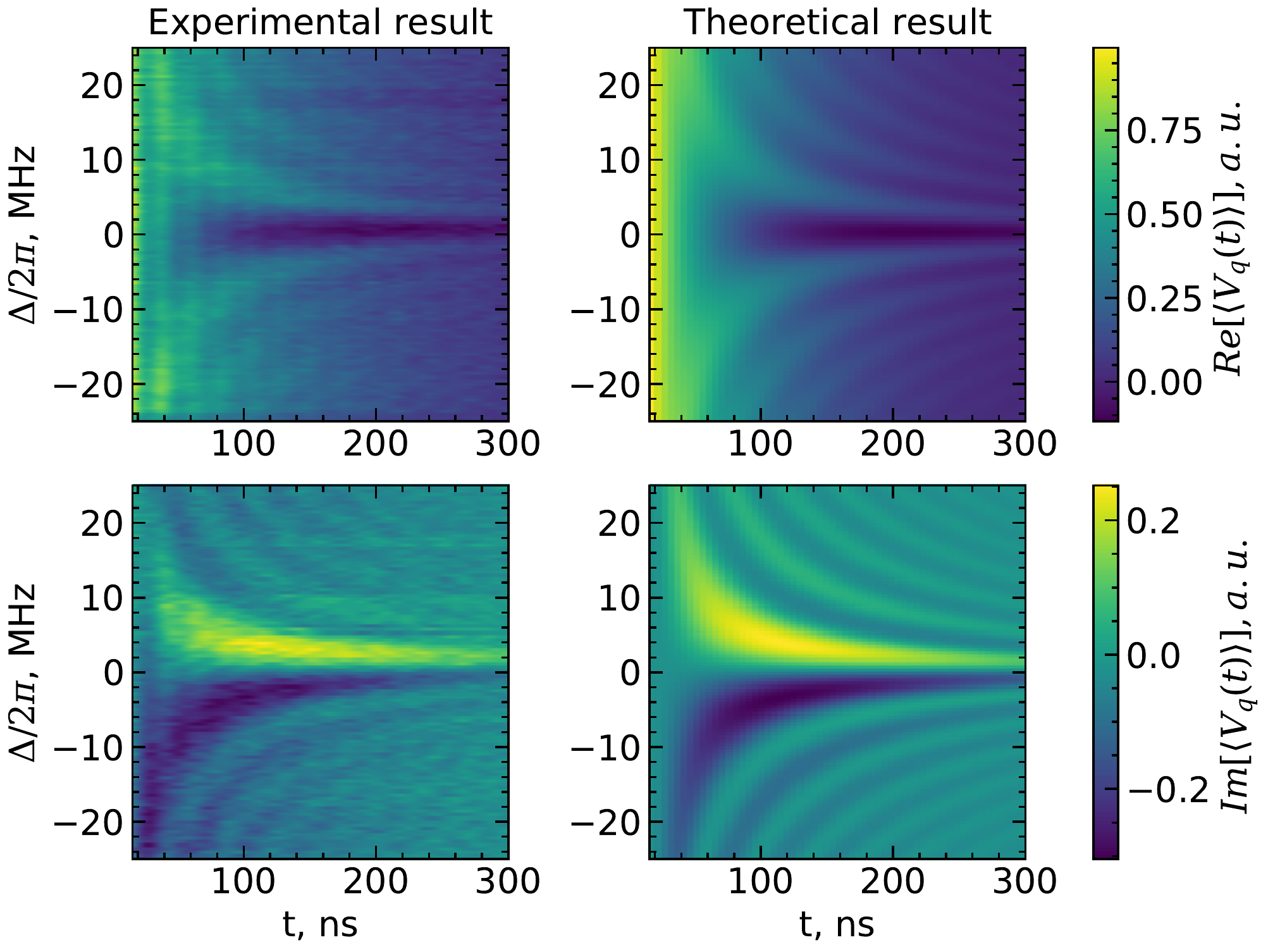}
\caption{\label{fig:2d_quantum_Im} Comparison of the detected field envelope $V_{q}(t)$ in real time $t$  with respect to detuning $\Delta$ between the photon carrier frequency $\omega_{c}$ and Probe qubit frequency $\omega_{p}$ with the analytical results, see Eq. (3), in case of scattering single-photon pulse emitted from the photon source, namely, Emitter on the Probe qubit. Both quadratures (real and imaginary) of the field are presented without removing the incoming single-photon pulse. The color bars for the experimental and theoretical quadratures are the same.}
\end{figure}

An approach to obtain an analytical expression for the scattered field is to consider the isolated system of two atoms interacting via a continuum of waveguide modes further utilizing Weisskopf \cite{scully1999quantum} approximation. Another approach is to include the interaction as a dissipation-like exchange term in the formalism of master equations \cite{Gardiner1985, Gardiner1994}. Omitting the propagation delay time, detected field consists of expectation values of lowering operators representing fields emitted by the Emitter and the field scattered by Probe:
\begin{equation}
    V_q(t) = A_e\sqrt{\gamma}\langle{\sigma^e_-(t)}\rangle + A_p\sqrt{\Gamma}\langle{\sigma^p_-(t)}\rangle,
\end{equation}
Here $A_{p} = \frac{A_{e}}{\sqrt{2}}$.
After some calculations (see App. A), one obtains a formula for the detected field:
\begin{equation}
V_{q}(t) \propto \sqrt{\gamma}\frac{e^{-\gamma_2 t}}{2}+\sqrt{\frac{\Gamma}{2}}\frac{\sqrt{\Gamma \gamma} \left(e^{-\gamma_2 t}-e^{-\Gamma_2 t+i \Delta  t}\right)}{\gamma-\Gamma+2i \Delta },
\label{eq: quantum_em}
\end{equation}
where $\Gamma_2$ stands for the dephasing rate of the Probe. The behaviour of the analytically derived $V_{q}(t)$, see right column of the Fig.~\ref{fig:2d_quantum_Im}, agrees fairly well with the measurement data. We clearly see that in the resonant case, the envelope does not oscillate, which is again a manifestation of not more than a single photon within the incoming pulse. However, for non-zero detunings the envelope oscillates with the detuning frequency $\Delta$, which is in agreement with our measurements. Experimental and analytical cases are fitted with the common scaling parameter in order to get joint color bar.
 
\subsection{\label{sec:level2}Coherent pulse scattering}
For quantum pulse, the scattered field part is fully defined (see. Eg. (3)) via radiative linewidths of Probe and Emitter. Therefore, it is possible to experimentally compare quantum case scattering results in the resonant condition of vacuum-photon carrier frequency and Emitter transition with the case of scattering of coherent pulse. The resonant case is the most convenient, since we avoid field oscillations, which are not connected with the differences in photon statistics. Hence, we fix the frequency and digitize scattered field $V_{cl}(t)$, with respect to amplitude $\Omega_{0}$ of the coherent exponentially modulated pulse.

The pulse is applied to the Probe atom, and its amplitude can be described by the effective Rabi frequency $\Omega(t) = \Omega_0e^{-\gamma_2t}$ \cite{Fischer}. 
The amplitude $\Omega_{0}$ of the classical pulse is a free parameter setting up the mean number of photons in the pulse. Experimentally measured field traces with respect to $\Omega_{0}$ are presented in the upper panel of Fig. 4. The incoming exponential pulse is digitally removed and the case is chosen to be resonant, thus we can resolve field dynamics connected with the photon statistics only. Since measurements are performed in the resonance condition, the only quadrature is plotted, since the other quadrature is trivial if incoming pulse is subtracted. We are interested in the case of single envelope oscillation, which is the closest classical analog of single-photon scattering. The scattered field depends on the ratio between $\Omega_0$ and $\Gamma$. If $\Omega_0 \gg \Gamma$, then the Probe undergoes  Rabi oscillations. We measure the scattered field for different values of $\Omega_0$, for future analysis we limit ourselves to the case of the $\Omega_{0}/2 \pi \ll 10$~MHz, which is the limit of $\Gamma \sim \Omega_{0}$. In the regime of $\Omega_0 \sim \Gamma$ the average number of photons in classical pulse may be less than one, and the scattering is more similar to the quantum case.

As in the previous section, we proceed to describe the scattered coherent pulse with an analytical model. To describe the scattering of coherent pulse with exponential envelope, we look for the solution of the master equation in Lindblad form for the Hamiltonian $H={\Omega_0}/{2} \cdot e^{-\frac{\gamma t}{2}} \sigma_{x}$. In the resonance case, the equations are:
\begin{equation}
    \begin{array}{r@{}l}
      \partial_{t}\langle \sigma_{y} \rangle = & -\Omega_0 \langle \sigma_{z} \rangle e^{-\frac{\gamma t}{2}}-\frac{\Gamma}{2}\langle \sigma_{y} \rangle\\
      \partial_{t} \langle \sigma_{z} \rangle = & -\Gamma(1+\langle \sigma_{z} \rangle)+\Omega_0 e^{-\frac{\gamma t}{2}} \langle \sigma_{y} \rangle
    \end{array}
\end{equation}
\begin{figure}[h]
\includegraphics[width=\linewidth]{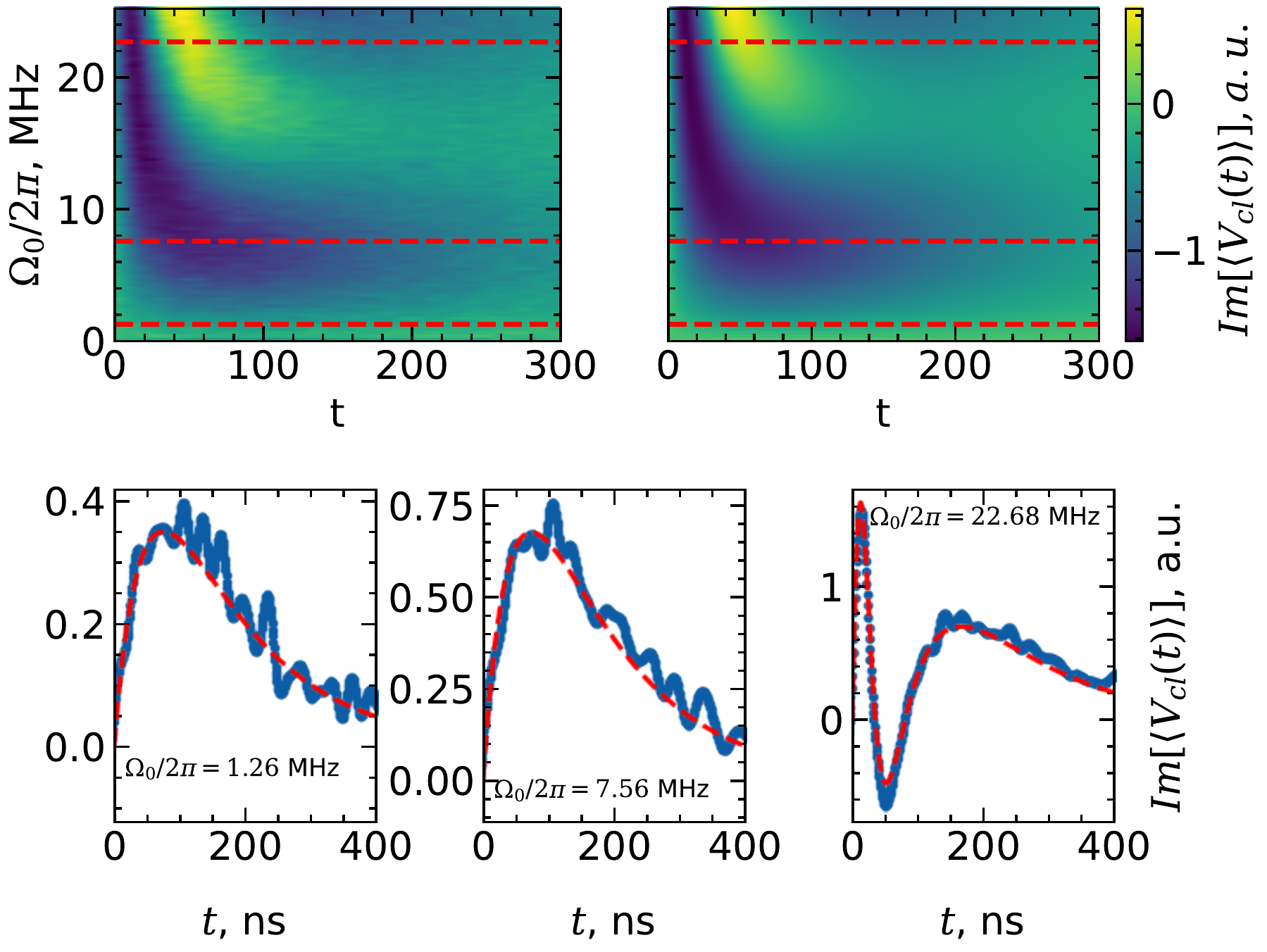}
\caption{\label{fig:2d_classical}Comparison of measured field envelopes $V_{cl}(t)$ (in resonance condition between the carrier frequency of the incoming exponentially modulated pulse and the Probe) in the time domain $t$ with respect to the strength $\Omega_{0}$ of driving pulse with theoretical results in case of classical field scattering. Incoming pulse is omitted. The color bars of the experimental and theoretical quadratures are the same. Red dashed lines on the color maps are the cuts with different $\Omega_{0}$. Corresponding traces are plotted at the bottom of the figure: dashed red line is theoretical fit, blue dots are the experimental traces.}
\end{figure}
The solution could be found as a series expansion in powers of $\Omega_0$, see Appendix B for a detailed procedure. In the first order, we get the linear in $\Omega_{0}$ part of answer:
\begin{equation}
   V_{cl}^{(1)}(t) = 
   \sqrt{\frac{\Gamma}{2}} \frac{\Omega_0 \left(e^{-\frac{\gamma t}{2}}-e^{-\frac{\Gamma t}{2}}\right)}{{\gamma} - {\Gamma}},
\label{eq:classical_correction1}
\end{equation}
Even orders have zero contribution, but, odd corrections are non-zero and decrease polynomially with respect to $\Omega_{0}$. The analytical solution is presented in the bottom panel of the Fig. 4. One can find a good correspondence of the experimental result with the analytical expressions. Scattered field envelope continuously transforms to the case of two atomic oscillations as the amplitude of the exciting signal increases. The only fit parameter is the prefactor, so the color bars are the same. Moreover, it is worth mentioning that leading order $V_{cl}^{(1)}$ fully matches the answer when the only photon is re-emitted by the Probe qubit, however, that question is not the not subject of the discussion, since it require generally more complex analytical treatment.

\section{Comparison of classical and quantum scattering}
Now we are ready to thoroughly compare both scattering cases and extrapolate some results to the wider range of radiative linewidth, since, experimentally we are limited in the experimentally accessible range of parameters. 
Since the only oscillation is present in the quantum case, one can potentially distinguish a non-classical state in the field which constitutes an input for the Probe, if the scattered field is measured. However, we note that for a very weak Rabi amplitude, the behaviour of the emitted field is also non-periodic. Thus we ask another, more complex question: whether one is able to distinguish between the classical and quantum cases when the number of photons in the classical pulse continuously approaches one, that is single oscillation case.

\begin{figure}[]
\includegraphics[width=\linewidth]{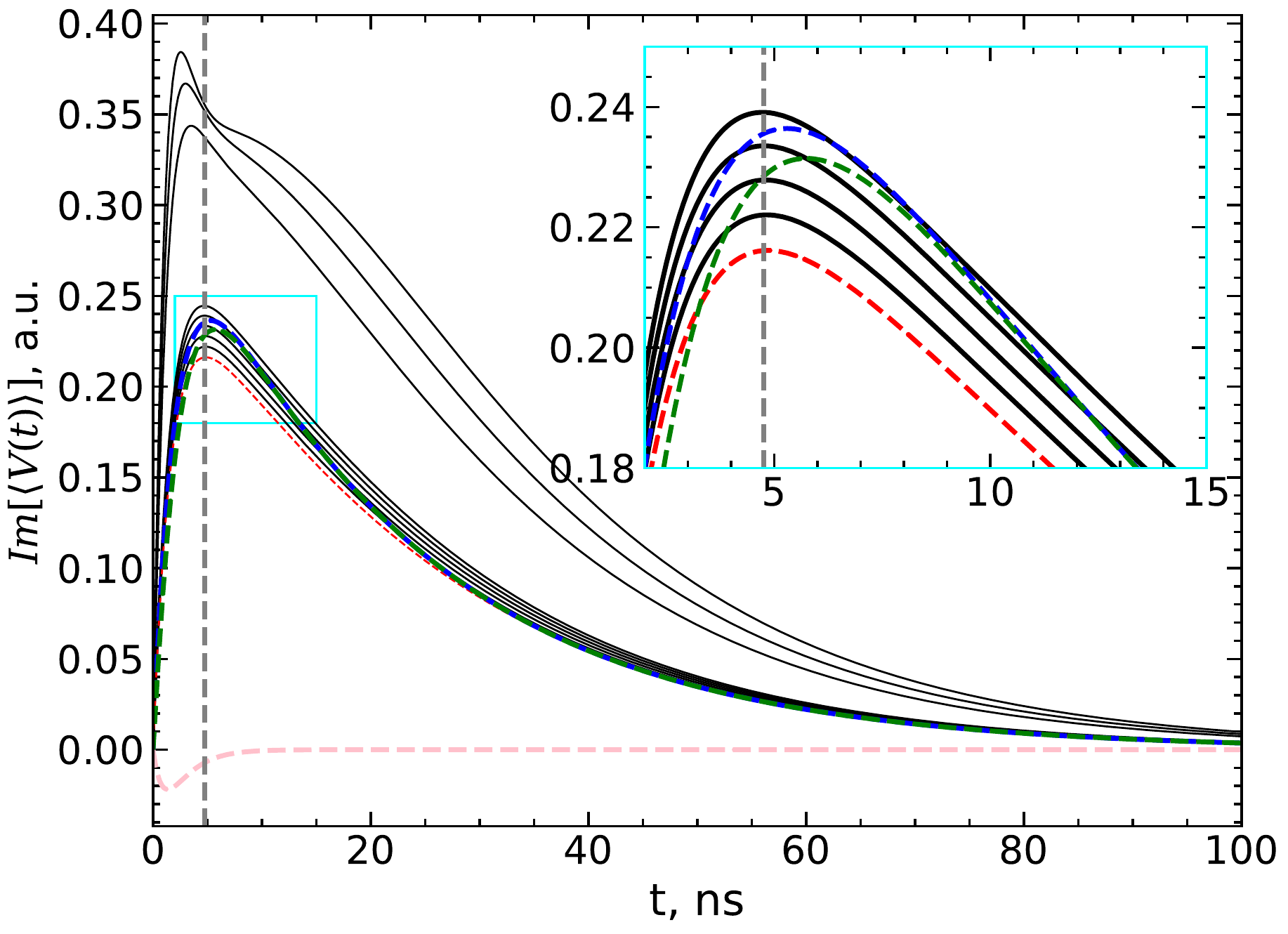}
\caption{\label{fig:compare_1d} Calculated scattered fields in both quantum and classical cases (the incoming exponential pulse is omitted in all cases).  Red dashed line: the numerical solution (matches full analytical solution) of the master equation with classical input pulse with $\Omega_0/2\pi = \sqrt{\gamma \Gamma}/2\pi= 0.3$~MHz. Black solid lines: the same for $\Omega_0/2\pi$ (0.31, 0.32, 0.33, 0.34, 0.35, 0.6, 0.7, 0.8)~MHz. Blue dashed line: the solution of the master equation with quantum input pulse, which fully matches (omitted for clarity) linear in $\Omega_0$ part of the solution for the incoming exponential classical pulse with $\Omega_0=\sqrt{\gamma \Gamma}$. Green dashed line: sum of two first non-zero contributions to the classical case solution with $\Omega_{0}/2\pi = 0.3$~MHz. Pink dashed line: the second non-zero (third order in $\Omega_0$) contribution of the classical solution, $\Omega_{0}/2\pi = 0.3$~MHz}
\end{figure}

Analyzing the results of calculations, we observe the difference between classical and quantum cases. In Fig.~\ref{fig:compare_1d}
we present the full analytical solution for classical and quantum pulses together with successive approximations of the first and third order. As stated before, the single-photon superposition radiated from the Emitter has the fixed amplitude defined in Eq. \eqref{eq: quantum_em}. Consequently, there is an optimal Rabi frequency $\Omega_*$ of classical pulse, for which the effect of this pulse on the Probe is the most similar to the effect of quantum superposition pulse. In other words, we could say that $\Omega_*$ is the effective ``classical'' amplitude of the quantum pulse. Comparing Eq.\eqref{eq: quantum_em} and Eq.\eqref{eq:classical_correction1} we find that $\Omega_0=\Omega_*\equiv \sqrt{\gamma\Gamma}$. Therefore, the quantum case solution Eq.\ref{eq: quantum_em} fully matches the first-order approximation of the classical case for $\Omega_0 = \Omega_*$. Importantly, $\Omega_{*}$ underestimates value of the amplitude closely matching the quantum case.

\begin{figure}[]
\includegraphics[width=\linewidth]{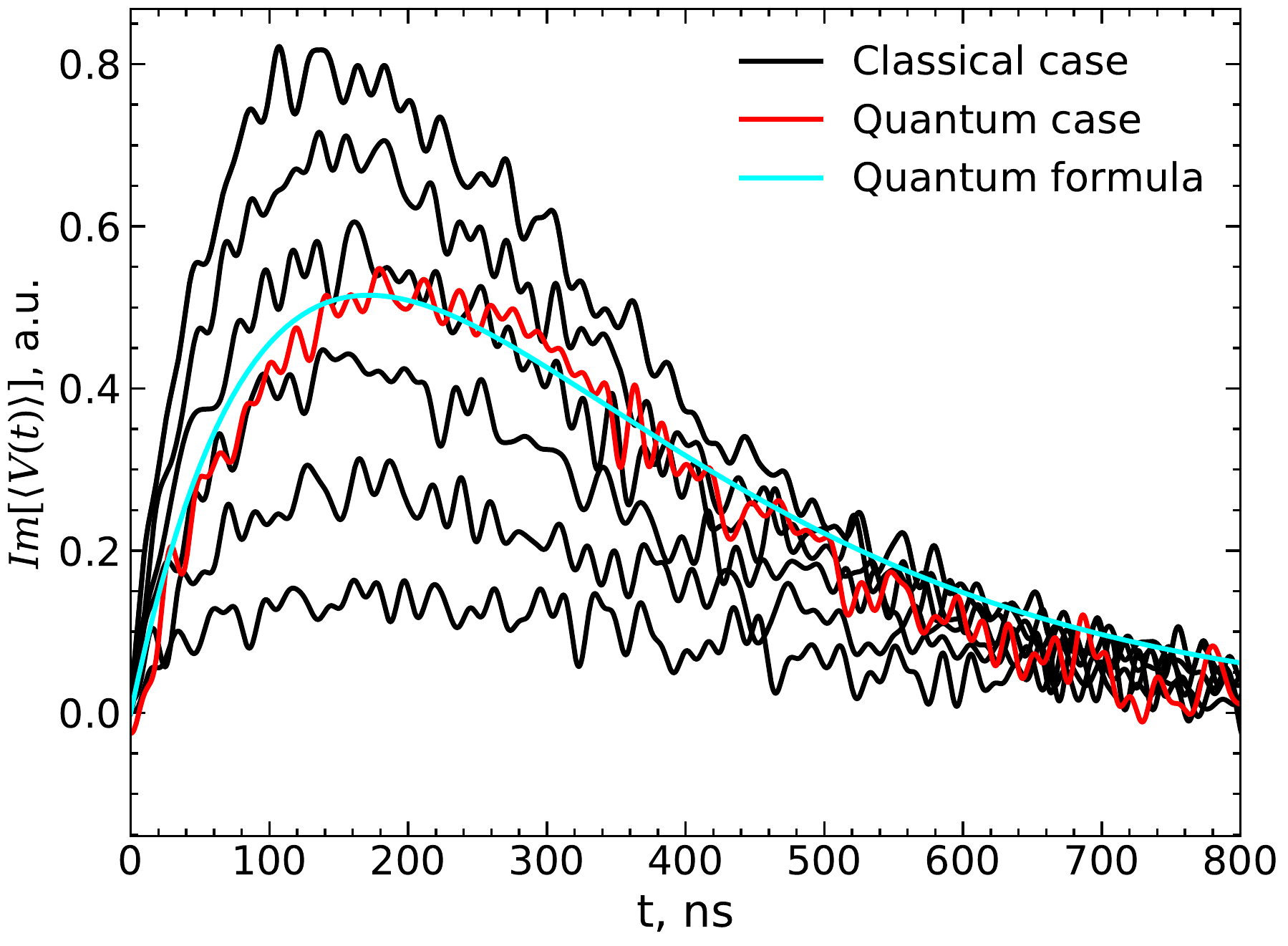}
\caption{\label{fig:resonance_comparison}Comparison of the experimentally obtained scattered fields in resonance case in both quantum and classical cases omitting incoming pulse. Scattering of the classical coherent pulse is provided with respect to the amplitude of the incoming pulse, corresponding $\Omega_{0}/2\pi$ frequencies are (0.27, 0.54, 0.81, 1.08, 1.36) MHz. Quantum case is fitted with Eq.\eqref{eq: quantum_em} (cyan solid line).}
\end{figure}

In the limit of small $\Omega_0 \ll \Gamma$ one should
expect vanishing, but finite, difference between classical and quantum cases for any ratio between $\gamma$ and $\Gamma$. For $\Omega_0 \ge \Gamma, \gamma$ the waveform changes and starts to behave oscillatory, which is a drastic difference of behaviour (the calculation is made for $\gamma/2\pi=0.09$~MHz, $\Gamma/2\pi=1$~MHz, giving the most demonstrative picture). Peak positions in two cases are shifted, but the shift is relatively small and it is an open question whether it could be measured with existing technical limitations of waveguide-QED setup since impedance mismatches and low signal-to-noise ratio could potentially suppress the effect.

We also demonstrate comparison of the experimentally measured field envelopes for both classical and quantum cases in Fig. \ref{fig:resonance_comparison}. It is impossible here to clearly distinguish the difference in envelopes. However, the closest classical signal could be determined in terms of Rabi amplitudes. $\Omega_*/2\pi = 1.02$~MHz results in the best approximation of the quantum answer to the classical one, which is slightly lower than $\sqrt{\Gamma\gamma}/2\pi \simeq 1.855$~MHz.

Next, in order to compare results of different calculations we use following integral distance norm between two functions:

\begin{equation}
    \epsilon=\sqrt{\int\limits_{0}^{\infty}dt\frac{(V_{cl}-V_{q})^{2}}{(V_{cl}+V_{q})^{2}}}
\end{equation}

\begin{figure}[]
\includegraphics[width=\linewidth]{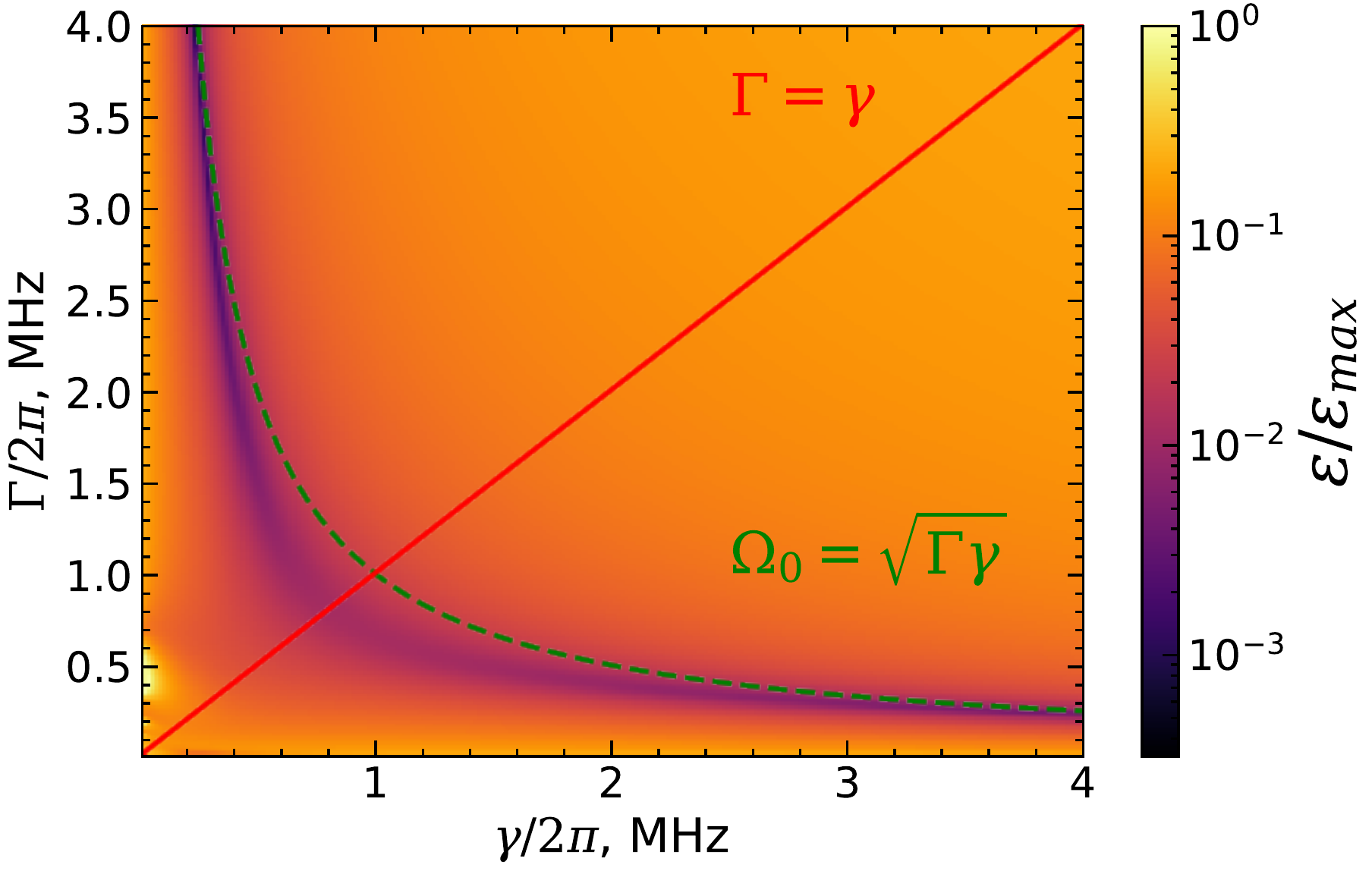}
\caption{\label{fig:compare_2d} Normalized dependence of the integral distance $\epsilon$ between calculated scattered fields in quantum and classical case on the relaxation decay rates of Emitter and Probe qubits. The amplitude of the classical signal is chosen to be $\Omega_{0}/2\pi=1$~MHz. The green line represents the harmonic mean of the relaxation rates of atoms, the red line is the mean line in the $\Gamma$, $\gamma$ coordinates.}
\end{figure}
\begin{figure}[]
\includegraphics[width=\linewidth]{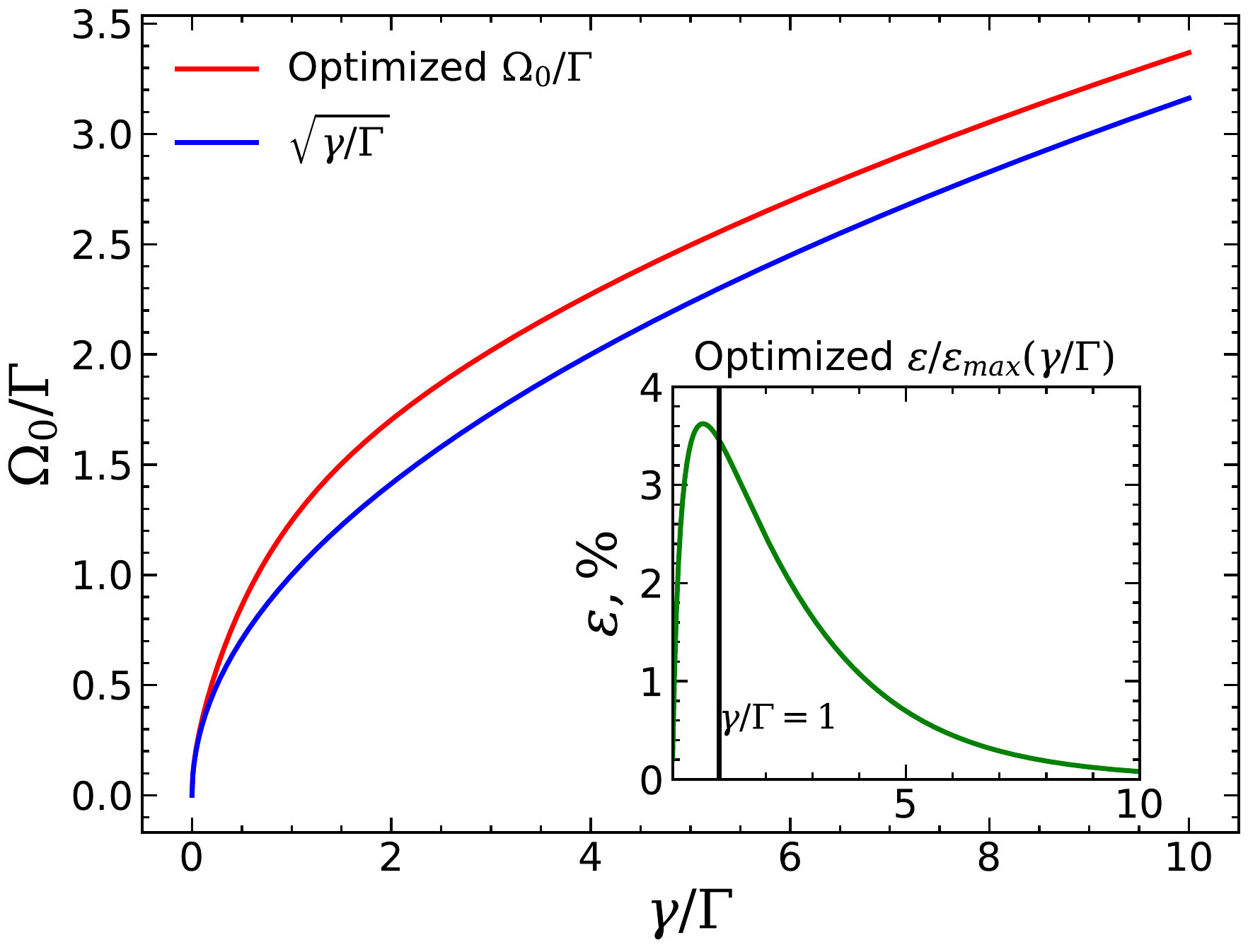}
\caption{\label{fig:compare_1d_epsilon} Numerically optimized $\Omega_{0}$ giving a minimum of the integral distance $\epsilon$ between the calculated scattered fields in both quantum and classical cases with respect to the ratio of decay times. The optimized value is compared with the harmonic mean of radiation rates divided by the relaxation rate of the Probe qubit. An inset picture describes the dependence of $\epsilon$ on the ratio of decay times. Each point corresponds to a minimum of the difference between classical and quantum answers.}
\end{figure}


which is a positive-valued function of decay constants and classical amplitudes. The normalized value of $\epsilon/\epsilon_{\text{max}}$, where $\epsilon_{\text{max}}=\max\limits_{\Gamma, \gamma}\epsilon$, is plotted in Fig.~\ref{fig:compare_2d} for fixed value of $\Omega_0$ as a function of $\Gamma \text{ and } \gamma$.  The dark area in Fig. \ref{fig:compare_2d} is the region where the quantum case has the closest correspondence with the classical case. We see that for $\gamma=\Gamma$ the discrepancy is maximal, and the optimal $\Omega_0$ is slightly less than $\sqrt{\gamma\Gamma}$. The difference between the two answers still not equals zero, and never equals zero assuming any non-zero choice of radiative rates, since higher order photon scattering processes are still present in the coherent signal. Therefore, we proceed with optimizing the $\epsilon$ numerically optimizing different values of $\Omega_0$.  In Fig. \ref{fig:compare_1d_epsilon} we show minimal value of normalized $\epsilon$ obtained as a function of $\frac{\Omega}{\Gamma}$ for different $\frac{\gamma}{\Gamma}$ ratios. One can see that the optimal distance between classical and quantum curves has a distinct maximum. That result closely relates to the case of equal decay rates (inset figure) where we see that optimized $\Omega$ is close to the half of mean harmonic. Another noticeable feature is that in the limit of $\gamma \gg \Gamma$ difference approaches zero, since the Probe is almost not excited by the incoming pulse. The spectral width of the pulse would be so great compared to the linewidth of the Probe qubit, that the incoming pulse does not really identify the scattering potential. Another case is $\gamma \ll \Gamma$ where qubit rapidly re-emits incoming photons so that they almost do not excite it. Therefore, even though there are many photons in the field, the second system relaxes too quickly, and the behaviour is not very different from the single photon case.

\section{Conclusion}
To conclude, we compared experimentally and theoretically the scattering of quantum single-photon superposition pulse and classical exponentially decaying pulse onto the Probe superconducting artificial atom. Experimentally we show that there is an optimal amplitude for the classical pulse, with which the behaviour of the envelope of the scattered field is most similar to the quantum case. Also, we experimentally show presence of the single field envelope oscillation in the quantum case scattering.
  We show that there is a non-vanishing difference between the envelopes of the scattered fields in the wide range of decay rates, which could be potentially observed in the experiments with the greater SNR. The results of the work could be potentially extrapolated to the scattering of pulses with other non-classical photon statistics or scattering on other systems e.g. an equidistant three level system.

\begin{acknowledgments}
The sample was fabricated using equipment of MIPT Shared Facilities Center. This research was supported by Russian Science Foundation, grant no. 21-42-00025.

\end{acknowledgments}

\nocite{*}
\bibliography{references}
\newpage
\appendix

\section{Solution for the quantum case}
\label{app:appa}
The cascaded setup being considered could be typically assumed as a closed type system (two TLS and bosonic bath), while neglecting external drive non-diagonal drive terms. Therefore, one could expand pure vectors of state in single photon basis, due to presence of integral of motion, namely, $\Sigma_{k}\langle a_{k}^{+}a_{k} \rangle+\Sigma_{i=1,2}\langle \sigma^{i}_{+}\sigma^{i}_{-}\rangle$. Thus, the problem of quantum scattering could be reviewed in terms of pure states formalism, expanded into single-photon basis, allowing one to solve system of ordinary differential equations, which can be modelled by means of Runge-Kutta methods or exactly solved proposing Markovian type approximations. On the other hand, we can not omit drive terms in the case of classical signal scattering, due to continuous filling of Fock states in coherent signal, thus the problem is more complex due to direct time-dependence of coefficients in the equations.
The system is described by the following Hamiltonian:

\begin{equation}    
    H =  H_{TLS} + H_{B} + H_{I1}  + H_{I2},
\end{equation}
where two-level systems have different energy splittings:
\begin{equation}    
H_{TLS} = \omega_e\hat{\sigma}_{1+}\hat{\sigma}_{1} +\omega\hat{\sigma}_{2+}\hat{\sigma}_{2},
\end{equation}
bosonic bath consists of left- and right-moving modes:
\begin{widetext}
\begin{equation}
H_B = \int_{0}^{+\infty} (dk) \omega(k)\hat{a}_R^\dag(k)\hat{a}_R(k)+  \int_{-\infty}^{0} (dk)\omega(k)\hat{a}_L^\dag(k)\hat{a}_L(k), H_{I1} = \int_{0}^{+\infty} (dk) g_1 \Big[\hat{a}_L^\dag(k)\hat{\sigma}_{1-} + \hat{a}_L(k)\hat{\sigma}_{1+} \Big],
\end{equation}
\end{widetext}

The first system interacts only with right-moving modes and the second system interacts with both left- and right-moving modes:
\begin{widetext}

\begin{equation}
\label{hi2}
H_{I2} = \int_{0}^{+\infty} (dk) g_2 \Big[e^{- i k r}\hat{a}_R^\dag(k)\hat{\sigma}_{2-} + e^{ i k r}\hat{a}_R(k)\hat{\sigma}_{2+}^\dag\Big]+
\int_{-\infty}^{0} (dk) g_2 \Big[e^{- i k r}\hat{a}_L^\dag(k)\hat{\sigma}_{2-} + e^{ i k r}\hat{a}_L(k)\hat{\sigma}_{2+}^\dag \Big].
\end{equation}
\end{widetext}

The system is prepared at $t=0$ in the state
\begin{equation}
    |\Psi\rangle = \frac{1}{\sqrt{2}}\Big(|\uparrow\rangle + e^{i\phi}|\downarrow\rangle \Big)|\downarrow\rangle|0\rangle.
\end{equation}
Note that this setup makes sense only for $r<0$ (the first system which is excited, interacts only with modes propagating to the left).

The number of photons never exceeds $1$ and at later times, the wave-function can be written as follows:
\begin{widetext}
\begin{equation}
|\Psi\rangle = \zeta(t)|\downarrow\rangle|\downarrow\rangle|0\rangle + \alpha_1(t) e^{-i\omega_e t}|\uparrow\rangle|\downarrow\rangle|0\rangle + \alpha_2(t) e^{-i\omega t}|\downarrow\rangle|\uparrow\rangle|0\rangle + \sum_{k}\beta_k(t)e^{-i\omega_k t}|\downarrow\rangle|\downarrow\rangle|k\rangle
\end{equation}
\end{widetext}

\begin{widetext}
\begin{equation}
\label{eq:ampl}
    \Dot{\zeta}(t) = 0,\quad\Dot{\alpha}_1 =-ig_1 \int_{k<0} (dk) e^{-i(\omega_k - \omega_e) t}\beta_k,\quad\Dot{\alpha}_2 =-ig_2 \int (dk)   e^{-i(\omega_k-\omega) t}e^{i k r}\beta_k
\end{equation}
\end{widetext}

and equation for field's amplitudes:
\begin{widetext}
\begin{equation}
    \Dot{\beta}_k = -ig_1 \Theta(-k)e^{i(\omega_k - \omega_e) t}\alpha_1 -ig_2 e^{i(\omega_k - \omega) t}e^{-i k r}\alpha_2.
\end{equation}
\end{widetext}

As expected, nothing happens with the $\ket{\downarrow}\ket{\downarrow}\ket{0}$ amplitude:
\begin{equation}
    \zeta(t) \equiv \frac{1}{\sqrt{2}}.
\end{equation}
The electric field's amplitudes can be found as follows:
\begin{widetext}
\begin{equation}
    \beta_k(t) = \int_{0}^{t}d\tau\Big[-ig_1 \Theta(-k)e^{i(\omega_k - \omega_e) \tau}\alpha_1(\tau) -ig_2 e^{i(\omega_k - \omega) \tau}e^{-i k r}\alpha_2(\tau)\Big].
\end{equation}
\end{widetext}

We can now substitute this expression to the Eq. (\ref{eq:ampl}) to get coupled integro-differential equations for 2LS amplitudes:
\begin{widetext}
\begin{equation}
\label{eq:ampl_id_1}
\Dot{\alpha}_1 =-\int_{k<0} (dk) e^{-i(\omega_k - \omega_e) t}\int_{0}^{t}d\tau\Big[g_1^2\Theta(-k)e^{i(\omega_k - \omega_e) \tau}\alpha_1(\tau)+g_1 g_2 e^{i(\omega_k - \omega) \tau}e^{-i k r}\alpha_2(\tau)\Big]
\end{equation}
\end{widetext}

and

\begin{widetext}
\begin{equation}
\label{eq:ampl_id_2}
\Dot{\alpha}_2 =-\int (dk) e^{-i(\omega_k-\omega) t}e^{i k r}\int_{0}^{t}d\tau\Big[g_1 g_2\Theta(-k)e^{i(\omega_k - \omega_e) \tau}\alpha_1(\tau) +g_2^2e^{i(\omega_k - \omega) \tau}e^{-i k r}\alpha_2(\tau)\Big].
\end{equation}
\end{widetext}

The Eqs (\ref{eq:ampl_id_1}), (\ref{eq:ampl_id_1}) can be further simplified by changing integration variable over each chiral branch:
\begin{equation}
\int_{k>0} (dk)\to \int_{0}^{\infty}\frac{d\omega}{2\pi} \frac{1}{|d\omega/dk|}.
\end{equation}

Next, we neglect momentum dependence of the group velocity: $d\omega/dk\to c$ and extend frequency integrations to infinity. This allows to derive the equations, which are local in time (using Weisskopf approximation of the integral over frequency) (with $\frac{\gamma}{2}=\frac{g_1^2}{c},\; \frac{\Gamma}{2}=\frac{2g_2^2}{c}$; we also neglect the time delay, $r/c\to 0$):

\begin{equation}
\begin{cases}
 \Dot{\alpha}_1(t) = -\frac{\gamma}{2}\alpha_1(t),\\
        \Dot{\alpha}_2(t) = -\frac{\Gamma}{2} \alpha_2(t)  -\frac{1}{2}\sqrt{\frac{\Gamma \gamma}{2}}  e^{-i(\omega_e - \omega)t}\alpha_1(t).
        \end{cases}
\end{equation}

Electric field can be found as follows
\begin{equation}
   V(x,t) \propto i \int (dk) e^{ikx-i\omega_k t} \beta_k(t) 
\end{equation}

which gives for the wave propagating to the left (we also put $r\to 0$):

\begin{equation}
    V(x,\tau=t-|x|/c) \propto e^{i\phi}\left[e^{-i\omega_e \tau}\alpha_1(\tau) + e^{-i\omega \tau}\alpha_2(\tau)\right]
\end{equation}
From there scattered part of the field in quantum case is given by
\begin{equation}
\label{eqn:q_analytical}
V_{q}(t) = a e^{i \phi} \sqrt{\frac{\Gamma}{2}}\frac{\sqrt{\gamma \Gamma} \left(e^{-\frac{\gamma t}{2}}-e^{-\frac{\Gamma t}{2}+i \Delta  t}\right)}{\gamma-\Gamma+2i \Delta }
\end{equation}

Where $\Delta = \omega_{e}-\omega$ is detuning between qubits, $\gamma$ - radiative relaxation rate of SPS, and $\Gamma$ - rate for Probe qubit. $\phi$ denotes delay of the wave package due to finite optical length, which can be experimentally fully removed. All signals have own affine transformation, which mixes field quadratures,  $a e^{i \phi}$ altering detected signal due to presence of experimental environment. Therefore, solutions on the quadratures of field should be odd and even with respect to detuning between qubits; calibration of the delay could be provided assuming resonance condition - one of the scattered field quadratures should be equal to zero. Therefore, we have only one free parameter for the subsequent analysis, namely, scaling factor, which accounts for overall attenuation and amplification in the experimental tract.

Assuming closed bath-TLS system allows one to obtain the solution in much more simple way using only vectors of state, which is not the case when modes of field are factorized inevitably leading to the density matrix approach.

However, analytical solution could be determined using equivalent master equation \cite{Gardiner1985}. Using input-output relation for the cascaded system and assuming Markovian approximation one is able to determine density matrix dynamics:
\begin{widetext}
\begin{eqnarray}
\partial_{t}\rho = -i\Big[\rho, \frac{\omega}{2}\sigma_{z}+\frac{\omega^{e}}{2}\sigma^{e}_{z}+\frac{i\sqrt{\gamma \Gamma}}{2}(\sigma^{e}_{+}\sigma_{-}-\sigma_{+}\sigma^{e}_{-})\Big]+\Big(\sqrt{\gamma}\sigma^{e}_{+}+\sqrt{\Gamma}\sigma_{+}\Big)\rho\Big(\sqrt{\gamma}\sigma^{e}_{-}+\sqrt{\Gamma}\sigma_{+}\Big)-\\
\nonumber
\frac{1}{2}\Big(\sqrt{\gamma}\sigma^{e}_{+}+\sqrt{\Gamma}\sigma_{+}\Big)\Big(\sqrt{\gamma}\sigma^{e}_{-}+\sqrt{\Gamma}\sigma_{+}\Big)\rho-\\
\nonumber
\frac{1}{2}\rho\Big(\sqrt{\gamma}\sigma^{e}_{+}+\sqrt{\Gamma}\sigma_{+}\Big)\Big(\sqrt{\gamma}\sigma^{e}_{-}+\sqrt{\Gamma}\sigma_{+}\Big)
\end{eqnarray}
\end{widetext}
With $\rho(0)=(\frac{|0\rangle+|1\rangle}{\sqrt{2}})(\frac{\langle0|+\langle1|}{\sqrt{2}})\otimes(|0\rangle\langle 0|)$
In the Heisenberg representation such system gives closed system of ODEs where cross-correlations between two Hilbert spaces couldn't be simply reduced like in effective mean field theories \cite{Mean-field} or similar Langevin equation systems\cite{Pogosov}. Nevertheless it is possible to find full analytical solution for the field in resonance case.

\newpage
\section{Derivation of classical case formulae}
\label{app:appb}

The very natural approach to compare results obtained after quantum scattering is to scatter classical field with same wave-form in time domain. One has to exponentially modulate classical pulse by means of arbitrary wave-form generator. Due to difference in photon statistics it is proposed that scattered field can possess different features. But, in the limit of infinitely small drive amplitude it is expected to obtain similar behaviour, since $\alpha$ in the coherent state would be excessively small. 

We will discuss solution of the closed system of equations of motion originating from the master equation in Linblad form.
In terms of detuning $\Delta$ between classical signal and Probe's qubit frequency and driving field force $\Omega$ one is able to formulate task in the rotating wave approximation:
\begin{equation}
H=\frac{\Delta}{2}\sigma_{z}+\frac{\Omega}{2} e^{-\frac{\gamma t}{2}} \sigma_{x}
\end{equation}
With radiative relaxation rate $\Gamma$ of the Probe (pure $\Gamma_1^{r}$) one is able to write following equations of motion using Heisenberg picture in resonance case:

\begin{equation}
    \begin{cases}
      \partial_{t}\langle \sigma_{x} \rangle = - \frac{\Gamma}{2}\langle \sigma_{x} \rangle\\
      \partial_{t}\langle \sigma_{y} \rangle = -\Omega \langle \sigma_{z} \rangle e^{-\frac{\gamma t}{2}}-\frac{\Gamma}{2}\langle \sigma_{y} \rangle\\
      \partial_{t} \langle \sigma_{z} \rangle = -\Gamma(1+\langle \sigma_{z} \rangle)+\Omega e^{-\frac{\gamma t}{2}} \langle \sigma_{y} \rangle
    \end{cases}
\end{equation}
Answer for $\langle \sigma_{-} \rangle$ contains only $\langle \sigma_{y} \rangle$ terms, which is nothing, but solution of the second order differential equation with exponential terms.
\begin{widetext}
\begin{equation}
   4 \Gamma \Omega  +e^{\frac{\gamma  t}{2}} \left ( 4 y''(t)+2 (\gamma +3 \Gamma ) y'(t)+\Gamma  (\gamma +2 \Gamma ) y(t) \right)+4 \Omega^{2} e^{-\frac{\gamma  t}{2}} y(t)=0
\end{equation}
\end{widetext}
Where $\langle\sigma_{y}\rangle(t)\equiv y(t)$, $y(0)=0$, $y'(0)=-\Omega$.
Closed system of the first order differential equations contains time dependent terms, however full analytical solution exists and fully matches numerical simulations given in the main text. We are interested in solutions depending only on $\Omega$, where $$\Omega \sim \gamma \sim \Gamma$$ we can try to provide perturbative analysis for $\langle \sigma_{-} \rangle$ dynamics in time domain.
Solutions for $\Omega=0$ are pure exponents.
Taking into account initial conditions one could write simple perturbation theory in order to find asymptotic solution in terms of functional series (higher-order terms for each order of partial sums should be $O(\Omega^{n+1})$:
\begin{equation}
    \begin{cases}
      \langle \sigma_{y} \rangle =\sigma_{y}^{(0)}+e^{-\frac{\Gamma t}{2}} \sum\limits_{n=1}^{+\infty} \Omega^{n}\sigma_{y}^{(n)}\\
      \langle \sigma_{z} \rangle =\sigma_{z}^{(0)}+ \sum\limits_{n=1}^{+\infty} \Omega^{n}\sigma_{z}^{(n)}
    \end{cases}
\end{equation}
Where $\langle \sigma_{y} \rangle (0) = 0$ and $\langle \sigma_{z} \rangle (0) = -1$.
Which leads to $\langle \sigma^{(n)}_{y} \rangle (0) = 0$, $\langle \sigma^{(n)}_{z} \rangle (0) = -1$ for each order of n, where $n=0$ denotes unperturbed solution ($\Omega=0$). For $n=0$ chains starts from $\langle \sigma^{(0)}_{z}\rangle=-1$, $\langle \sigma^{(0)}_{y}\rangle=0 $. Recurrence relations for the given set of series are, provided condition of $n>0$:
\begin{equation}
    \begin{cases}
      \partial_{t}\langle \sigma_{y}^{(n)} \rangle \Omega^{n} =-\Omega e^{-\frac{\Gamma t}{2}-\frac{\gamma t }{2}}\left(\Omega^{n-1}\langle \sigma^{(n-1)}_{z} \rangle \right)\\
      \partial_{t}\langle \sigma_{z}^{(n)} \rangle \Omega^{n} = -\Omega^{n}\Gamma \langle \sigma_{z}^{(n)}\rangle +\Omega e^{\frac{\Gamma t}{2}-\frac{\gamma t }{2}}\left(\Omega^{n-1} \langle \sigma^{(n-1)}_{y} \rangle \right)
    \end{cases}
\end{equation}
Any possible correction could be de termined.
The first leading correction to the answer linear in $\Omega$:
\begin{equation}
\label{eqn:classical_correction1}
   V^{(1)}_{cl}(t)=\sqrt{\frac{\Gamma}{2}}\frac{\langle \sigma_{y} \rangle}{2} = \sqrt{\frac{\Gamma}{2}}\frac{\Omega \left(e^{-\frac{\gamma t}{2}}-e^{-\frac{\Gamma t}{2}}\right)}{\gamma -\Gamma}
\end{equation}
It is worth mentioning that such formula fully matches answer for the leading order in $\Omega$ for the scattered field in case of re-emitting not more than one photon. It could be easily derived utilizing Moller \cite{Fischer} scattering amplitudes with non-Hermitian Hamiltonian.

The second order correction in $\Omega$ is:

\begin{widetext}
\begin{equation}
   V^{(2)}_{cl}(t) =  \sqrt{\frac{\Gamma}{2}} \frac{2 \Omega^{3} e^{-t \left(\frac{3 \gamma }{2}+\Gamma \right)} \left(\left(\gamma-\Gamma\right)^{2} e^{\frac{1}{2} t (3 \gamma +\Gamma )}+\left(-3 \gamma ^2-2 \gamma 
   \Gamma +\Gamma ^2\right) e^{\frac{1}{2} t (\gamma +\Gamma )}+\gamma  (3 \gamma -\Gamma ) e^{\gamma  t}+\gamma  (\gamma +\Gamma ) e^{\Gamma  t}\right)}{\gamma  (\gamma -\Gamma )^2 (3 \gamma
   -\Gamma ) (\gamma +\Gamma )}
\end{equation}
\end{widetext}

Other corrections are too complex to reflect them explicitly, but non-zero and polynomically decreasing.

\newpage





\end{document}